\documentclass{iaus}
\usepackage{graphicx}

\title[ Globular Cluster Systems] %% give here short title %%
{Clues to Globular Cluster Evolution from Multiwavelength Observations of Extragalactic Systems}

\author[Kundu et al.]   %% give here short author list %%
{Arunav Kundu$^1$,
Thomas J. Maccarone$^2$, \and Stephen E. Zepf$^1$}

\affiliation{$^1$Department of Physics \& Astronomy, Michigan State University, East Lansing, MI 48824, USA
 \\ email: {\tt akundu@pa.msu.edu, zepf@pa.msu.edu} \\[\affilskip]
$^2$School of Physics \& Astronomy, University of Southampton, Southampton, UK SO17 1BJ \\email: {\tt tjm@phys.soton.ac.uk}}

\pubyear{2007}
\volume{246}  %% insert here IAU Symposium No.
\pagerange{}
\jname{Dynamical Evolution of Dense Stellar Systems}
\editors{E. Vesperini, M. Giersz, A. Sills, eds.}
\begin{document}

\maketitle

\begin{abstract}
We present a study of the globular cluster (GC) systems of nearby elliptical and S0 galaxies at a variety of wavelengths from the X-ray to the infrared. Our analysis shows that roughly half of the low mass X-ray binaries (LMXBs), that are the luminous tracers of accreting neutron star or black hole systems, are in clusters. There is a surprisingly strong correlation between the LMXB frequency and the metallicity of the GCs, with metal-rich GCs hosting three times as many LMXBs as metal-poor ones, and no convincing evidence of a correlation with GC age so far. In some of the galaxies the LMXB formation rate varies with GC color even within the red peak of the typical bimodal cluster color distribution, providing some of the strongest evidence to date that there are metallicity variations within the metal-rich GC peak as is expected in a hierarchical galaxy formation scenario. We also note that any analysis of subtler variations in GC color distributions must carefully account for both statistical and systematic errors. We caution that some published GC correlations, such as the apparent 'blue-tilt' or mass-metallicity effect might not have a physical origin and may be caused by systematic observational biases.

\end{abstract}

\firstsection % if your document starts with a section,
\section{Introduction}

High resolution Chandra X-ray images of nearby ellipticals and S0s have resolved large numbers of point sources,  confirming a long-standing suggestion that the hard X-ray emission in many of these galaxies is predominantly from X-ray binaries. In early type galaxies most of the bright, L$_X$$\gtrsim$10$^{37}$ erg s$^{-1}$ sources seen in typical Chandra observations must be low mass X-ray binaries, binary systems comprising a neutron star or black hole accreting via Roche lobe overflow from a low mass companion, since they generally have stellar populations that are at least a few Gyrs old.  

An important characteristic of LMXBs is that they are disproportionately abundant in globular clusters. Even though GCs account for $\lesssim$0.1\% of the stellar mass in the Galaxy, they harbor about 10\% of the L$_X$$\gtrsim$10$^{36}$ erg s$^{-1}$ LMXBs , indicating  a probability of hosting a LMXB that is at least two orders of magnitude larger than for field stars. This has long been attributed to efficient formation of LMXBs in clusters due to dynamical interactions in the core. Early type galaxies are ideal for studies of the LMXB-GC link as they are particularly abundant in globular clusters. The identification of LMXBs with these simple stellar systems that have well defined properties such as metallicity and age provides a unique opportunity to probe the effects of these parameters on LMXB formation and evolution.

\begin{figure}[b]
% \vspace*{-2.0 cm}
\begin{center}
 \includegraphics[width=5in]{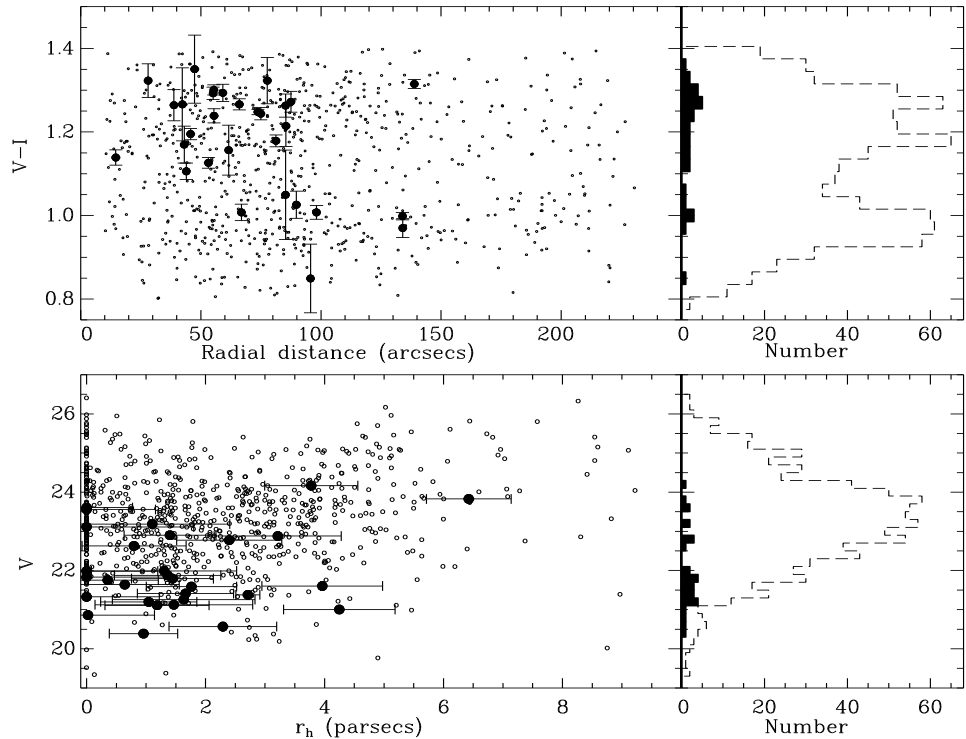} 
% \vspace*{-1.0 cm}
 \caption{Top: The V-I colors of GCs vs. distance from the center of NGC 4472 and GC color distribution. LMXB-GC matches are indicated by filled circles/bins.  Bottom: V magnitude of globular clusters vs. half light radius and the globular cluster luminosity function. LMXBs are preferentially located in the brightest, most metal-rich GCs. There is a weak anti-correlation with GC half-light radius and no obvious correlation with galactocentric distance. Each of these broad correlations (or lack thereof) have been confirmed in other galaxies. }
   \label{fig1}
\end{center}
\end{figure}

\section{The Effect of GC Environment on LMXB Formation \& Evolution  }

Fig 1 plots the colors, magnitudes, half-light radii, and galactocentric distances of globular clusters identified in HST-WFPC2 images of NGC 4472 (\cite[Kundu, Maccarone \& Zepf 2002]{2002ApJ...574L...5K} [KMZ02]), the brightest elliptical in the Virgo cluster. The well known Gaussian globular cluster luminosity function and the bimodal globular cluster color (metallicity) distribution are apparent. The large symbols indicate the GCs that host LMXBs. LMXBs are preferentially found in the most luminous and red (metal-rich) GCs. Statistical tests (KMZ02) reveal a marginal tendency of LMXBs to favor GCs with smaller half light radius, and no convincing correlation with galactocentric distance. These correlations have subsequently been confirmed in other galaxies (\cite[Kim et al. 2006]{2006ApJ...647..276K}; \cite[Kundu, Maccarone, \& Zepf 2007]{2007ApJ...662..525K} [KMZ07]; \cite[Sivakoff et al. 2007] {2007ApJ...660.1246S}). The presence of bright LMXBs in $\approx$4\% of GCs and the association of $\approx$40\% of LMXBs with GCs in NGC 4472 is also typical of the values in other early type galaxies.

\begin{figure}[t]
% \vspace*{-2.0 cm}
\begin{center}
 \includegraphics[width=5in]{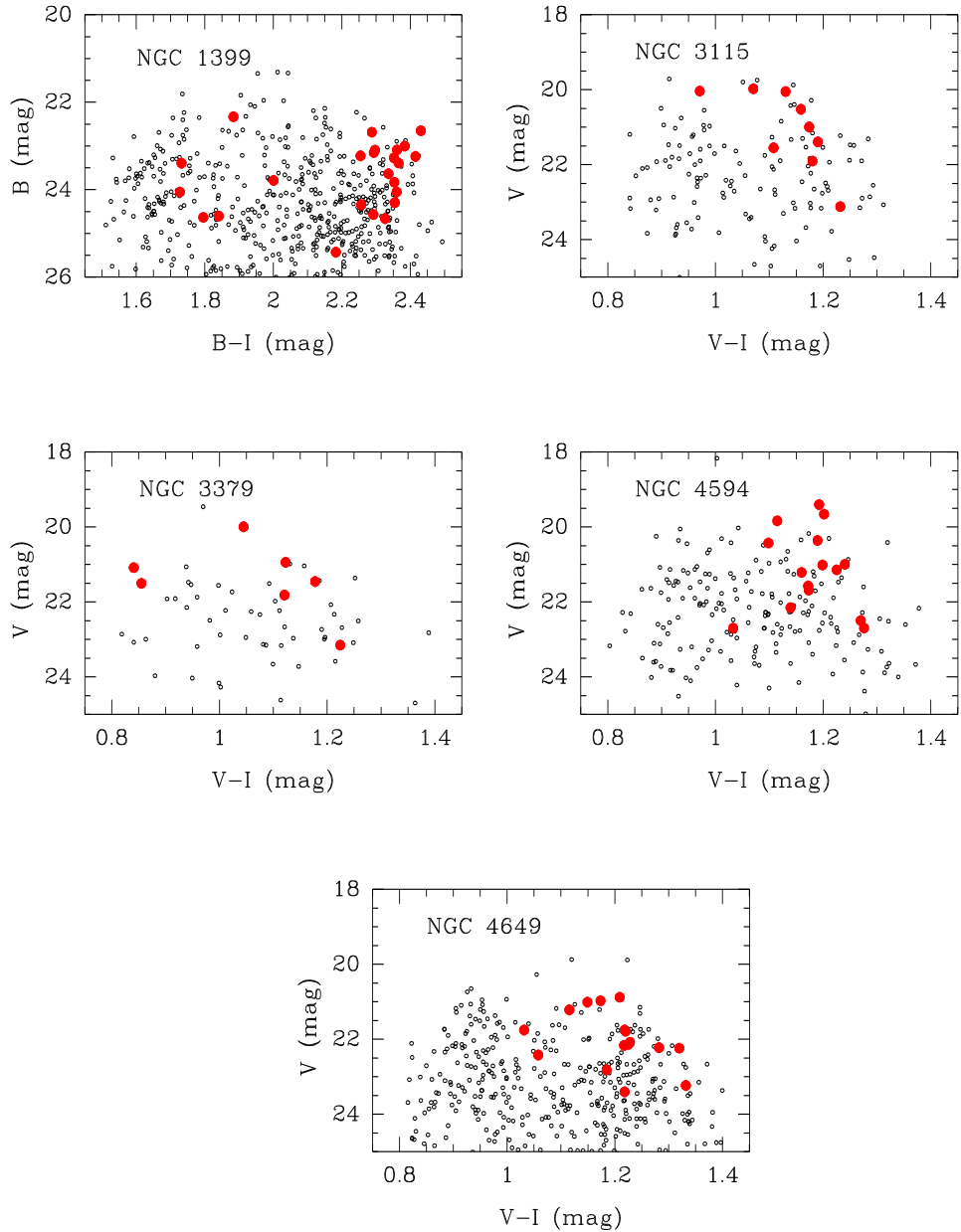} 
% \vspace*{-1.0 cm}
 \caption{Color-magnitude diagrams for globular cluster candidates in five elliptical galaxies with known bimodal cluster metallicity distributions. Filled points represent clusters with LMXB counterparts. LMXBs are clearly found preferentially in luminous (high mass) and red (metal-rich) globular clusters. There is a clear enhancement of LMXBs in the reddest GCs in the metal-rich GCs of NGC 1399 providing some of the strongest evidence to date that there is metallicity substructure within the red GCs in this giant elliptical galaxy.}
   \label{fig1}
\end{center}
\end{figure}

	The underlying reasons for these correlations provide a window into both the dynamics of GCs and the physics of LMXBs. Since luminous clusters are known to be denser than less luminous ones and obviously have more stars, the higher rate of LMXB formation in these clusters due to the consequently higher dynamical interaction rate is not surprising. One the other hand there are no obvious dynamical reasons for the three times larger rate of LMXBs in the red, metal-rich, globular clusters as compared to the metal-poor ones. LMXBs are similarly found preferentially in the metal-rich Galactic GCs, but in the past this was often dismissed as the consequence of the location of these clusters in the bulge, where the dynamical evolution of GCs may be accelerated. The lack of correlation of the LMXBs with galactocentric distance in the large GC samples of ellipticals (KMZ02; KMZ07) argues against this possibility and establishes a primary correlation with GC color.

	It has also been suggested that enhanced rate of LMXBs in metal-rich GCs may be because of the younger ages of these clusters. However infrared imaging, which yields more accurate constraints on the metallicities of GCs and helps  disentangle the age-metallicity degeneracy, has established that the optical colors of globular cluster systems that have clearly bimodal distributions are primarily a tracer of metallicity (\cite[Kundu \& Zepf 2007]{2007ApJ...660L.109K}) and LMXBs are indeed preferentially located in metal-rich GCs (\cite[Kundu et al. 2003]{2003ApJ...589L..81K}; \cite[Hempel et al. 2007]{2007ApJ...661..768H}). Fig 2 shows the color magnitude distributions of five galaxies with confirmed bimodal color distributions and the factor of three enhancement with metallicity.

A very interesting feature of the NGC 1399 distribution is that not only is there a larger fraction of LMXBs in the metal-rich sub-population of clusters, but in fact LMXBs are preferentially located in the most metal-rich GCs. This is the most convincing evidence to date that there is metallicity structure within the metal-rich peak of GCs, as is expected from hierarchical models of galaxy and globular cluster system formation. It is not clear if the reason that this feature is obvious only in NGC 1399 is because of the location of this galaxy at the center of the Fornax cluster which leads to a particularly efficient enrichment history, or because of the larger color baseline of this data set. Although the Sivakoff et al. (2007) analysis is broadly in agreement about the metallicity trend and finds a linear increase in the fraction of LMXBs in GCs with metallicity, it suggests that the rate of LMXBs in the reddest, and consequently most metal-rich GCs actually drops. We note that LMXBs are found preferentially in the brightest globular clusters, and if the GC sample in Sivakoff et al. (2007) were to be restricted to similar luminosities this apparent discrepancy would disappear. In other words the reddest GCs in the Sivakoff sample are the faint ones which have scattered to these colors due to observational uncertainties and are not representative of the true underlying color/metallicity of the GCs.

	There have been some theoretical attempts to explain the metallicity effect. \cite[Maccarone, Kundu, \& Zepf (2004)]{2004ApJ...606..430M} suggest that irradiation of the donor star by the LMXB is key. Higher metallicity objects can dissipate this energy through line cooling while a wind is generated in low metallicity star, thus lowering the LMXB lifetime. \cite[Ivanova (2006)]{2006ApJ...636..979I} on the other hand suggests that the absence of an outer convective layer in solar mass metal-poor stars limits magnetic braking and the formation of mass transferring LMXBs.

Various groups have attempted to derive the dependence of the observed LMXB rate on the globular cluster metallicity and mass by assuming a specific M/L ratio and color-metallicity correlation (\cite[Jordan et al. 2004]{2004ApJ...613..279J}; \cite[Smits et al. 2006]{2006A&A...458..477S}; Sivakoff et al. 2007). These studies generally agree that the mass dependence is roughly linear and the metallicity dependence is approximately Z$^{0.25}$. Jordan et al. (2004) and Sivakoff (2007) further attempt to link the probability of finding a LMXB in a GC to the rate of stellar interactions in the core of a GC. However, this requires measuring the core radii of GCs which is a small fraction of even a HST ACS pixel for these galaxies. Thus the core radius is derived by extrapolation of other measured GC properties. Smits et al. (2006) show that the Jordan et al. (2004) results are as expected if there is no information about the core radius of the clusters in their sample. While the core radii of extragalactic GCs are a challenge for the present data sets, the half-light radii can indeed be measured and its effect on other measured GC parameters must be accounted for carefully. Some of the consequences of not doing so are outlined next.

\section{ 'Blue-Tilts', Mass-Metallicity Trends and other Correlations in the Properties of Halo Globular Clusters }

	Many recent HST-based studies of globular cluster systems have reported the discovery of color-metallicity trends in the blue, metal-poor sub-population of globular clusters (\cite[Strader et al. 2006]{2006AJ....132.2333S}; \cite[Mieske et al. 2006]{2006ApJ...653..193M}; \cite[Harris et al. 2006]{2006ApJ...636...90H}). This small trend of brighter blue clusters appearing a few hundredths of a magnitude redder per magnitude of brightness has been dubbed the 'blue-tilt', and is viewed as evidence of a mass-metallicity trend. This has been interpreted as evidence of self-enrichment of the more massive clusters either because they started out with dark matter halos that have since been stripped, or because they formed in large gas clouds. This effect is larger in more massive galaxies, and more apparent when the color magnitude diagram is plotted using the redder of the two filters under consideration for the magnitude axis (Mieske et al. 2006).

	We have reanalyzed some of these data sets and find that there is a small, but measurable, mass-radius relationship for GCs in these galaxies (Kundu \& Zepf 2007b in preparation). Since the blue GCs are on average larger, this trend is better defined for the metal-poor clusters. Studies of extragalactic cluster systems typically assume either a uniform aperture correction or attempt to fit the profile within a small aperture in order to minimize the uncertainties introduced by the galaxy background. We show in Kundu \& Zepf (2007b) that the blue tilt is likely the consequence of photometric bias introduced by the effect of the mass-radius relationship on such photometric techniques.

	It is also important to note that the uncertainties in color and magnitude are not independent, and hence not orthogonal on a color magnitude diagram. In fact when the redder of the two filters is used for the magnitude axis in a color-magnitude diagram the uncertainties are parallel to the direction of the 'blue-tilt' and hence amplify the effect. Conversely the uncertainties tend to negate the 'blue-tilt' when the  blue filter is plotted on the magnitude axis of a color magnitude plot, thus explaining the filter effect. Moreover, the underlying galaxy light is known to be redder for more massive galaxy. This explains the apparent variation of the strength of the 'blue-tilt' with galaxy mass.

	Finally we note that various studies have suggested that there is a small color metallicity trend in the mean colors of the metal-poor globular clusters with host galaxy mass (e.g. Strader et al. 2006). We find a trend of larger mean sizes of blue GCs with larger galaxy mass, which may cause this effect. Thus, within the observational uncertainties the properties of metal-poor halo clusters appear to be remarkably uniform everywhere.

\end{document}